\documentclass[a4paper,11pt]{article}
\usepackage{latexsym}
\usepackage{cite}
\usepackage{epsfig}
\usepackage{amssymb}
\author{Parviz Goodarzi \footnote{parviz.goodarzi@abru.ac.ir} \space and \space H. Mohseni Sadjadi \footnote{mohsenisad@ut.ac.ir}
\\ {\small Department of Science, Ayatollah Boroujerdi University}
\\ {\small Department of Physics, University of Tehran,}
\\ {\small P. O. B. 14395-547, Tehran 14399-55961, Iran}}
\title {Reheating in a modified teleparallel model of inflation}
\begin{document}
\maketitle
\begin{abstract}
We study the cosmological inflation and reheating in a teleparallel model of gravity.
Reheating is assumed to be due to the decay of a scalar field to radiation during its rapid oscillation.
By using cosmological perturbations during inflation,
and subsequent evolutions of the Universe, we calculate the reheating temperature
as a function of the spectral index and the power spectrum.
\end{abstract}
\section{Introduction}

To solve problems arisen in the cosmological standard model such as the flatness,
the absence of monopoles, the isotropy and homogeneity in large scale and so on, the inflation model was introduced \cite{guth,inflaton1,Liddle,inflation2,inflation3,inflation4,Martin}.
Creation of small density inhomogeneity from quantum fluctuations in the early Universe is one of the
most important predictions of the cosmic inflation \cite{mukhanov1}.
In the standard inflation model, based on Einstein's theory of general relativity, a canonical scalar field (inflaton) during its slow roll drives the cosmic acceleration. Afterward, the reheating era begins, during which the inflaton begins a coherent oscillation and
generates radiation \cite{Linde2,Linde3,Alb,Sh,kof11,kof12}. At the end of reheating era, the Universe becomes radiation dominated. The temperature at this time is dubbed as the reheating temperature $T_{rh}$. Constraints from the big bang nucleosynthesis (BBN), light elements abundance, and large scale structure and CMB put the lower bound $4MeV$ on reheating temperature\cite{han}. In addition, as the reheating occurs after inflation, the reheating temperature must be less than the GUT energy scale which is around $10^{16}GeV$.

Recently, the theory of gravity in the teleparallel framework \cite{e1,e2,e3,e4} has attracted more attention\cite{tel1,tel2,tel3,tel4,tel5,tel6,tel7,tel8,tel9}. This is due to the capacity of this model to describe the late time acceleration of the Universe \cite{teld1,teld2,teld3,teld4,teld5,teld6,teld7,teld8,teld9,teld10}, as well as the inflation in the early Universe\cite{telinf1,telinf2}. In the teleparallel model, the curvatureless Weitzenbock connections are used instead of the torsionless Levi-Civita connections employed in the Einstein theory of gravity. Similar to the well-known extension of Einstein-Hilbert action to modified gravity ( $f(R)$ model), the modified teleparallel gravity ($f(T)$ model) is an extension of the teleparallel model \cite{teld1,teld2}. Scalar and tensor perturbations in teleparallel gravity were studied in \cite{telinf3,telinf4}. Power spectrum and spectral index for scalar and tensor modes in $f(T)$ gravity have been calculated in \cite{telinf4}.

In this paper, inspired by the above-mentioned models, we will consider inflation in the modified teleparallel model. In a pure teleparallel model, it is not clear how the Universe is warming up after the inflation, and how particles are created. So we consider also a scalar field which decays to ultra-relativistic particles after the inflation, in a period of its rapid oscillation \cite{Sh,kof11,kof12,good1, good2}. By studying the evolution of the universe, we compute the reheating temperature as a function of the observable parameters such as the spectral index and the power spectrum derived from Planck 2018 data \cite{planck}.

The scheme of the paper is as follows:
In the second section, first we introduce the model and after some preliminaries, we briefly review inflation and cosmological perturbations in the power law modified teleparallel cosmology.
In the third section, which is the main part of the paper, by studying the evolution of the Universe, and by using the results of the second section, the reheating temperature is calculated.
We use units $\hbar=c=k_B=1$ through the paper.

\section{Model introduction and preliminaries}

To study the inflation and the subsequent reheating, we consider the modified teleparallel gravity with a canonical scalar field and radiation described by the following action \cite{teld9}
\begin{equation}\label{1}
S=\int{\bigg[{1\over16\pi G}f(T)+
{1\over2}\partial_\mu\varphi\partial^{\mu} \varphi- V(\varphi)+\pounds_r\bigg]}ed^4x ,
\end{equation}
where $e=det({e^A}_\mu)=\sqrt{-g}$, $M_P=\sqrt{{1\slash(8\pi G)}}=2.4\times 10^{18}GeV$ is the reduced Planck mass,
$\pounds_r$ is the radiation`s lagrangian density, $T$ is the torsion scalar which is constructed by contraction of the torsion tensor
\begin{equation}\label{2}
T={1\over4}T_\rho^{\,\ \mu\nu}T^\rho_{\,\ \mu\nu}-{1\over2}T_{\,\,\,\,\,\rho}^{\mu \nu}T^\rho_{\,\,\mu\nu}-T_{\,\,\rho\mu}^{\rho}T^{\nu\,\,\mu}_{\,\,\nu}.
\end{equation}
The torsion tensor is given by
\begin{equation}\label{3}
T^\rho_{\,\ \mu\nu}=\Gamma^\rho_{\mu\nu}-\Gamma^\rho_{\nu\mu}=e_A^\rho(\partial_\mu e^A_\nu-\partial_\nu e^A_\mu).
\end{equation}
Note that the radiation component becomes only relevant after inflation, i.e. in the reheating era.

The teleparallel is formulated with the veirbein fields $e_A^\mu$, in terms of which, the metric is given by
$g_{\mu\nu}(x)=\eta^{AB} e_{A}^{\mu} e_{B}^{\nu} $. For the spatially flat FLRW (Friedmann-Lema\^{i}tre-Robertson-Walker) metric $ds^2=dt^2-a^2(t)\delta_{ij}dx^idx^j$, the evolution of the scale factor, $a(t)$, is given by the Friedmann equations
\begin{eqnarray}\label{4}
&&H^2={1\over 3M_P^2}(\rho_T+\rho_\varphi+\rho_{r}), \nonumber \\
&&\dot{H}=-{1\over 2M_P^2}(\rho_T+\rho_\varphi+\rho_{r}+P_T+P_\varphi+P_{r}),
\end{eqnarray}
where $ H=\dot{a}/a$ is the Hubble parameter, a "dot" is differentiation with respect to cosmic time $t$,
and prime denotes differentiation with respect to the scalar field $\varphi$. The torsion scalar is $T=-6H^2$, and the energy density and the pressure of the scalar field are
\begin{eqnarray}\label{5}
\rho_\varphi={\dot{\varphi}^2\over 2}+V(\varphi),\nonumber \\
P_\varphi={\dot{\varphi}^2\over 2}-V(\varphi).
\end{eqnarray}
$\rho_T$ and $P_T$ are determined by
\begin{eqnarray}\label{6}
&&\rho_T={M_P^2\over2}\bigg(2Tf_{,T}-f-T \bigg) ,\nonumber \\
&&P_T=-{M_P^2\over2}\bigg(-8\dot{H}Tf_{,TT}+(2T-4\dot{H})f_{,T}-f+4\dot{H}-T \bigg).
\end{eqnarray}
The equation of state (EoS) parameter is defined by $w:=\gamma-1:=\frac{P}{\rho}$.
The components involved in (\ref{4}) satisfy the continuity equations
\begin{eqnarray}\label{7}
\dot{\rho_T}+3H(\rho_T+P_T)=0 ,\nonumber \\
\dot{\rho_{\varphi}}+3H(\rho_{\varphi}+P_{\varphi})=-Q_0,\nonumber \\
\dot{\rho_r}+4H\rho_r=Q_0,
\end{eqnarray}
where $Q_0$ is the interaction term which becomes operative in the reheating era.
In the inflationary epoch $\varphi$ satisfies
\begin{equation}\label{8}
\ddot{\varphi}+3H\dot{\varphi}+V'(\varphi)=0.
\end{equation}

\subsection{Inflation }
The inflationary phase is specified by $\ddot{a}>0$. This is equivalent to $\dot{H}+H^2>0$. In terms of the slow roll parameter $\varepsilon_1$ defined by
\begin{equation}\label{10}
\varepsilon_1\equiv-{\dot{H}\over H^2},
\end{equation}
the inflation condition is $\varepsilon<1$. We adopt a power law modified teleparallel model \cite{telinf4}
\begin{equation}\label{s1}
f(T)=CT^{1+\delta},
\end{equation}
where $\delta$ is a nonnegative integer number, and $C=\frac{1}{M^{2\delta}}$, where $M$ is a constant with mass dimension . Hence
\begin{equation}\label{11}
\rho_T=-3M_P^2H^2\bigg((1+2\delta)CT^\delta-1 \bigg).
\end{equation}
From relation (\ref{4}) we have
\begin{equation}\label{12}
H^2={T^{-\delta}\over 3M_P^2C(1+2\delta)}\rho_\varphi.
\end{equation}
Inserting  $T=-6H^2$ in (\ref{12}) gives
\begin{equation}\label{12.1}
T^\delta=\bigg(-{2\over M_P^2C(1+2\delta)}\rho_\varphi\bigg)^{\delta\over\delta+1}.
\end{equation}
Therefore
\begin{equation}\label{12.2}
H^2={2^{{-\delta\over\delta+1}}\over3}\bigg({\rho_\varphi\over 3(-1)^\delta M_P^2C(1+2\delta)}\bigg)^{1\over1+\delta}.
\end{equation}
For $P_T$, we have
\begin{equation}\label{13}
P_T=-{M_P^2\over2}\bigg(-4C(1+\delta)T^\delta\bigg(3H^2+\dot{H}(1+2\delta)\bigg)-CT^{1+\delta}+4\dot{H}+6H^2\bigg).
\end{equation}
Using (\ref{13}), one can rewrite the equation (\ref{4}) as
\begin{equation}\label{14}
\dot{H}=-{3\over2}{H^2\over(1+\delta)}-{T^{-\delta}\over 2M_P^2C(1+\delta)(1+2\delta)}P_\varphi.
\end{equation}
Therefore the slow roll parameter becomes
\begin{equation}\label{15}
\varepsilon_1=-{\dot{H}\over H^2}={3\over2}{1\over 1+\delta}\bigg(1+{P_\varphi\over\rho_\varphi} \bigg).
\end{equation}
For $\delta=0$, this relation reduces to its well known form of the standard inflation model, i.e.  $\varepsilon_1=-{\dot{H}\over H^2}={3\over2}\bigg(1+{P_\varphi\over\rho_\varphi} \bigg)$.
${P_\varphi\over\rho_\varphi}\simeq -1$, which is equivalent to the slow roll condition, guarantees $\ddot{a}>0$.
The slow roll condition implies:
\begin{eqnarray}\label{16}
\rho_\varphi\approx V(\varphi)\nonumber \\
3H\dot{\varphi}\approx V'(\varphi).
\end{eqnarray}
Therefore (\ref{15}) reduces to
\begin{equation}\label{17}
\varepsilon_1\approx M_P^2{C(1+2\delta)T^\delta\over 1+\delta}\bigg({V'(\varphi)\over V(\varphi)}\bigg)^2.
\end{equation}
In the standard inflation model ($\delta=0$), (\ref{17}) becomes $\varepsilon_1\approx {M_P^2\over 2}\bigg({V'(\varphi)\over V(\varphi)}\bigg)^2 $.
By substituting $T^\delta$ from (\ref{12.1}) in (\ref{17}), we obtain
\begin{equation}\label{17.1}
\varepsilon_1\approx-{\bigg(-M_P^2C(\delta+{1\over2})\bigg)^{\big({1\over1+\delta}\big)}\over(1+\delta)}{V^{\prime}(\varphi)^2\over V(\varphi)^{\big({\delta+2\over\delta+1}\big)}}.
\end{equation}
For our future computation we need also to the other slow roll parameter $\varepsilon_2$ which is obtained as
\begin{equation}\label{17.2}
\varepsilon_2\equiv {\dot{\varepsilon_1}\over H\varepsilon_1}\approx {\dot{T}\delta\over TH}-{2V(\varphi)\over3H^2}
\bigg({V^{\prime\prime}(\varphi)\over V(\varphi)}-
\bigg({V^{\prime}(\varphi)\over V(\varphi)}\bigg)^2\bigg).
\end{equation}
This can be rewritten
\begin{equation}\label{17.3}
\varepsilon_2\approx -2M_P^2C(1+2\delta)T^\delta
\bigg({V^{\prime\prime}(\varphi)\over V(\varphi)}-{2+\delta\over 2+2\delta}\bigg({V^{\prime}(\varphi)\over V(\varphi)}\bigg)^2\bigg),
\end{equation}
which, as a function of $\varepsilon_1$, is
\begin{equation}\label{17.4}
\varepsilon_2\approx 2\varepsilon_1
\bigg((2+\delta)-2(1+\delta)\bigg({V^{\prime\prime}(\varphi)V(\varphi)\over V^{\prime}(\varphi)^2}\bigg)\bigg).
\end{equation}
For the power law potential $V(\varphi)=\lambda \varphi^n$, $\varepsilon_1$ and $\varepsilon_2$ become:
\begin{eqnarray}\label{17.5}
\varepsilon_1&&\approx-{n^2(-M_P^2C(\delta+{1\over2})\lambda^\delta)^{{1\over1+\delta}}\over 1+\delta }
\varphi^{{n\delta\over\delta+1}+2}, \nonumber \\
\varepsilon_2&&\approx 2\varepsilon_1{2+2\delta-n\delta\over n}.
\end{eqnarray}

The number of e-folds from $t_{*}$ in the inflation era, until the end of the inflation, $t_{end}$, is given by
\begin{equation}\label{18}
\mathcal{N}_I=\int_{t_{*}}^{t_{end}}Hdt\approx\int_{\varphi_{*}}^{\varphi_{end}}{H\over\dot{\varphi}}d\varphi
\approx-\int_{\varphi_{*}}^{\varphi_{end}}{3H^2\over V^\prime(\varphi)}d\varphi.
\end{equation}
By using (\ref{12.2}), (\ref{18}) becomes
\begin{equation}\label{18.1}
\mathcal{N}_I\approx{2^{-\delta\over1+\delta}\over(M_P^2(-1)^\delta C(1+2\delta))
^{1\over1+\delta}}\int_{\varphi_{end}}^{\varphi_{*}}{V(\varphi)^{1\over1+\delta}\over V^\prime(\varphi)}d\varphi.
\end{equation}
\subsection{ Cosmological perturbations}
Cosmological perturbations of this model have been considered in \cite{telinf3,telinf4}, where the spectral index and power spectrum have been calculated. Here we briefly review and list the main results. The spectral index is \cite{telinf4}
\begin{equation}\label{23}
n_s-1\approx -2\varepsilon_1-\varepsilon_2.
\end{equation}
This relation gives the spectral index as a function of the slow roll parameters.
The power spectrum is given by
\begin{equation}\label{24}
\mathcal{P}_s(k_0)\approx {H^2\over 8\pi^2M_P^2c_s^3\varepsilon_1}\bigg|_{c_sk=aH}.
\end{equation}
This relation must be evaluated at the horizon crossing for which $c_sk=aH$. The sound speed, $c_s$, is
\begin{equation}\label{25}
c_s^2={f_{,T}\over f_{,T}-12H^2f_{,TT}}.
\end{equation}
For $f(T)=CT^{1+\delta}$, the sound speed becomes
\begin{equation}\label{26}
c_s^2={1\over 1+2\delta},
\end{equation}
which is a constant. For  $\delta>0$ this speed is upper luminal.
By using equations (\ref{17}) and (\ref{25}) we obtain
\begin{equation}\label{27}
\mathcal{P}_s(k_0)\approx -{T^{-2\delta}(1+\delta)\over 12\pi^2C^2M_P^6\sqrt{1+2\delta}}{V(\varphi)^3\over V^{\prime}(\varphi)^2}\bigg|_{c_sk=aH}.
\end{equation}
By using (\ref{12.1}) and (\ref{17.1}), we obtain $\mathcal{P}_s(k_0)$, at the horizon crossing, as
\begin{equation}\label{30}
\mathcal{P}_s(k_0)\approx -{(-M_P^2C(\delta+{1\over2}))^{2\delta\over1+\delta}(1+\delta)\over12\pi^2C^2M_P^6\sqrt{1+2\delta}}{V(\varphi)^{3+\delta\over1+\delta}\over V^\prime(\varphi)^2}\bigg|_{c_sk=aH}.
\end{equation}
The spectral index as a function of the scalar field potential is obtained as
\begin{equation}\label{30.1}
1-n_s\approx 2\varepsilon_1
\bigg((3+\delta)-2(1+\delta)\bigg({V^{\prime\prime}(\varphi)V(\varphi)\over V^{\prime}(\varphi)^2}\bigg)\bigg)
\end{equation}

\section{Reheating temperature}
  After giving a glimpse of the model and a summary of the required equations, hereafter we begin our main discussion, and  try to obtain the reheating temperature. This reheating is assumed to be  due to the inflaton decay to ultra-relativistic particles during its coherent rapid oscillation. We follow the method used in  \cite{Mielczarek1,good1}, and consider the following distinct epochs:  1--The inflation era, from $t_\star$ (exit time of a pivot scale from the Hubble radius) until the end of inflation $t_{end}$. In this time the main density is $\rho_\varphi$. 2-- Rapid oscillation period, from $t_{end}$ until $t_{reh}$, where the (thermal) radiation becomes dominant. 3-- From $t_{reh}$ until the recombination $t_{rec}$  and finally from $t_{rec}$ until the present time $t_0$. The e-folds number from horizon crossing until now is then given by
\begin{eqnarray}\label{31}
\mathcal{N}=\ln{({a_0\over a_\star})}&=&\ln{({a_0\over a_{rec}})}+\ln{({a_{rec}\over a_{reh}})}+\ln{({a_{reh}\over a_{end}})}+\ln{({a_{end}\over a_{\star}})}\\\nonumber
&:=&\mathcal{N}_{4}+\mathcal{N}_{3}+\mathcal{N}_{2}+\mathcal{N}_{1}
\end{eqnarray}
In the following subsections we will derive $\mathcal{N}$ for each period.

It is important to note that during rapid oscillation if one adopts the perturbative approach, then the radiation (ultra-relativistic particles)becomes dominant at $a_{reh}$, i.e. thermalization occurs when $a=a_{RD}$, that is when the radiation dominates. But if we consider preheating, the perturbative approach fails and  $a_{RD}\neq a_{reh}$. This is due to the fact that we may have a large number of non-thermal produced particles shortly after the beginning of rapid oscillation. This issue will be discussed in the second subsection.
\subsection{Slow roll inflation}

During the slow roll inflation, the Hubble parameter varies slowly $\frac{\dot{H}}{H^2}\ll 1$, and the energy density of the scalar field is dominant.
by relation (\ref{18.1}) the number of e-folds for the potential $V(\varphi)=\lambda \varphi^n $ is
\begin{equation}\label{32}
\mathcal{N}_1\approx{(\delta+1)(-M_P^2C(\delta+{1\over2})\lambda^\delta)^{\big(-{1\over1+\delta}\big)}\over n(2+2\delta-n\delta)}\varphi_\star ^{\big({2+2\delta-n\delta\over1+\delta}\big)},
\end{equation}
where $\varphi(t_\star)=\varphi_\star$. For the special case $V(\varphi)={1\over 2}m^2\varphi^2$ we have
\begin{equation}\label{33}
\mathcal{N}_1\approx-{(\delta+1)\over4(-C(1+2\delta))^{1\over1+\delta}(m^2)^{\delta\over1+\delta}}\bigg({\varphi_\star\over M_P}\bigg)^{2\over1+\delta}.
\end{equation}
Using (\ref{30.1}), we write the spectral index for power law potential as
\begin{equation}\label{34}
1-n_s\approx-{2n(2+2\delta+n-n\delta)(-M_P^2C(\delta+{1\over2})\lambda^\delta)^{\big({1\over1+\delta}\big)}\over(1+\delta)}
\varphi_\star^{\big({n\delta\over\delta+1}-2\big)}.
\end{equation}
From (\ref{33}) and (\ref{34}) we obtain
\begin{equation}\label{35}
\mathcal{N}_1\approx \bigg({2+2\delta+n-n\delta\over 2+2\delta-n\delta}\bigg){1\over 1-n_s},
\end{equation}
which for $V(\varphi)={1\over 2}m^2\varphi^2$ reduces to $\mathcal{N}_I\approx {1/(1-n_s)}$.
The relation (\ref{30}), for the power law potential, becomes
\begin{equation}\label{36}
\mathcal{P}_s(k_0)\approx -{(-M_P^2C(\delta+{1\over2}))^{2\delta\over1+\delta}(1+\delta)\lambda^{\big({1-\delta\over1+\delta}\big)}
\over12\pi^2n^2C^2M_P^6\sqrt{1+2\delta}}\varphi_\star^{\big[n\big({1-\delta\over1+\delta}\big)+2\big]},
\end{equation}
where $\varphi_\star$ is the scalar field at the horizon crossing.
The Hubble parameter at the horizon crossing is
\begin{equation}\label{37}
H^2\approx 8\pi^2M_P^2c_s^3\varepsilon_1\mathcal{P}_s(k_0)\bigg|_{c_sk=aH}
\approx4\pi^2M_P^2{n(1-n_s)(1+2\delta)^{-3\over2}\over(2+2\delta+n-n\delta)}\mathcal{P}_s(k_0)\bigg|_{c_sk=aH}.
\end{equation}
At the end of inflation we have $\varepsilon_1\approx 1$, therefore
\begin{equation}\label{38}
\varphi_{end}\approx {\bigg({-M_P^2C\lambda^\delta(\delta+{1\over2})}\bigg)}^{\big({1\over2+2\delta-n\delta}\big)}
{\bigg(-{n^2\over1+\delta}\bigg)}^{\big({1+\delta\over 2+2\delta-n\delta}\big)},
\end{equation}
and
\begin{equation}\label{38.1}
\rho_{end}\approx\lambda \varphi_{end}^n\approx{\bigg({-M_P^2C(\delta+{1\over2})}\bigg)}^{\big({n\over2+2\delta-n\delta}\big)}
{\bigg(-{n^2\over1+\delta}\bigg)}^{\big({n(1+\delta)\over 2+2\delta-n\delta}\big)}\lambda^{\big({2(1+\delta)\over2+2\delta-n\delta}\big)}.
\end{equation}

\subsection{Rapid oscillation}

Based on the CMB anisotropies measurements and the relative abundances of light elements,
we know that at the beginning of the big-bang nucleosynthesis (BBN) the Universe was in thermal equilibrium in a radiation dominated era with a temperature satisfying $T_{reh}>T_{BBN}$.
This thermalization occured in a period after inflation which we call reheating epoch which ended at $a=a_{reh}$. We begin this part in the context of the original pertubative approach \cite{inflaton1,Mielczarek1,Alb}, then point out briefly the required modifications in the presence of preheating.

In the reheating era, the Universe was composed of a rapid oscillating scalar field ($\varphi$), and particles to which $\varphi$ decayed . The radiation (ultra-relativistic particles)stress tensor is
\begin{equation}\label{19.1}
T^{\mu\nu}_r=(\rho_r+P_r)u^{\mu}u^{\nu}+P_r g^{\mu\nu}.
\end{equation}
where $u^{\mu}$ is the four velocity of the radiation. We take the energy transfer as \cite{Alb}
\begin{equation}\label{20}
 Q_{\mu}=-\Gamma u^{\nu}\partial_{\mu}\varphi \partial_{\nu}\varphi,
\end{equation}
therefore
\begin{equation}\label{21}
\nabla_{\mu}T^{\mu\nu}_r=Q^{\nu}.
\end{equation}
Similarly the continuity equation for $\varphi$ gives
\begin{equation}\label{22.0}
\nabla_{\mu}T^{\mu\nu}_{\varphi}=-Q^{\nu}.
\end{equation}
In a comoving frame, (\ref{20}) becomes $Q_{0}=\Gamma \dot{\varphi}^2$ and the equation (\ref{21}) reduces to
\begin{equation}\label{22}
\dot\rho_r +3H(\rho_r+P_r)=\Gamma \dot{\varphi}^2.
\end{equation}
In the same way, (\ref{22.0}) reduces to
\begin{equation}\label{22.1}
\dot\rho_{\varphi} +3H(\rho_{\varphi}+P_{\varphi})=-\Gamma \dot{\varphi}^2.
\end{equation}
The inflaton's equation of motion is then
\begin{equation}\label{ref2}
\ddot{\varphi}+(3H+\Gamma)\dot{\varphi}+V'(\varphi)=0.
\end{equation}
During its rapid coherent oscillation, the inflaton decayed to ultra-relativistic bosonic and fermionic particles.
Following \cite{Alb,kof,kolb}, we have specified this decay by inserting the phenomenological friction term $\Gamma \dot{\varphi}^2$ into the main equation of motion.
A fundamental derivation of this term requires a fuller understanding of the nature of inflaton and its interactions. If
we consider a three-legged interaction of the form $S_{int.}=\int \sqrt{-g}d^4x(-\sigma\varphi \chi^2-h\varphi \bar{\psi}\psi)$, where $\chi$ and $\psi$ are bosonic and fermionic fields, the decay rate in tree level is derived as $\Gamma=\frac{\sigma^2}{8\pi m}+\frac{h^2m}{8\pi}$ \cite{inflaton1}. This multiplies the scalar field solution by an exponential decay factor $e^{- \gamma \Gamma t/2}$ \cite{inflaton1}, which is the same as the effect of the friction term in (\ref{ref2}).

During the inflation the scalar field decreases very slowly, and then after the slow roll, starts a rapid oscillation, through which generates relativistic particles. This quasiperiodic oscillation was discussed in \cite{Sh,kof11,kof12} and is described by
\begin{equation}\label{39}
\varphi=\Phi(t)\sin\bigg(\int\omega(t) dt\bigg).
\end{equation}
The scalar field EoS parameter, $w_\varphi$, is derived as \cite{Sh,good1,good2}
\begin{eqnarray}\label{40}
\gamma&=&w_\varphi+1={<\rho_\varphi+P_\varphi>\over <\rho_\varphi>}={<\dot{\varphi}^2>\over <\rho_\varphi>}\nonumber\\
&=&{2<\rho_\varphi-V(\varphi)>\over V(\Phi)}\nonumber\\
&=&2{\int_{-\Phi}^{\phi}\sqrt{\rho_\varphi-V(\varphi)}d\varphi\over{\int_{-\Phi}^{\phi}{1\over\sqrt{\rho_\varphi-V(\varphi)}}d\varphi}} \nonumber\\
&=&{2n\over n+2}.
\end{eqnarray}
The average is taken over an oscillation (For more details see \cite{Sh,good1,good2}).

For the power law potential, one can show that $<\dot{\varphi}^2>\simeq \gamma\rho_{\varphi}$ \cite{good1}. Hence (\ref{22.1}) may be rewritten as
\begin{equation}\label{ref1}
{\dot{\rho}}_{\varphi}+3H\gamma\rho_{\varphi}+\gamma\Gamma\rho_{\varphi}=0
\end{equation}
In the beginning of oscillations the scalar field is the dominant component of the Universe and in addition we take $\Gamma\ll 3H$ \cite{kolb,good1}, but later, as $H$ decreases, this approximation fails and the third
term in (\ref{ref1}) gains the same order of magnitude as the second term. For $\Gamma\ll 3H$, by ignoring the interaction term we approximate
\begin{eqnarray}\label{41}
\rho_\varphi= \rho_{\varphi,0}a^{-3\gamma}(t).
\end{eqnarray}
From (\ref{12.2}) and (\ref{41}) we deduce
\begin{eqnarray}\label{42}
a(t)\propto t^{{2(\delta+1)\over3\gamma}}.
\end{eqnarray}
In this era the Hubble parameter is approximated by
\begin{eqnarray}\label{43}
H\approx {2(\delta+1)\over 3\gamma t}.
\end{eqnarray}
By putting this back into (\ref{ref1}), the scalar field energy density, in the next approximation, is derived as
\begin{eqnarray}
\rho_\varphi&=&\rho_{end}\left(\frac{t_{end}}{t}\right)^{2(\delta+1)}e^{-\Gamma\gamma(t-t_{end})}\nonumber \\
\rho_\varphi&=&\rho_{end}\left(\frac{a_{end}}{a}\right)^{3\gamma}e^{-\gamma\Gamma(t-t_{end})},
\end{eqnarray}
where $t_{end}$ is the end of the slow roll, i.e.  when the oscillation begins.
The term $\left(\frac{t_{end}}{t}\right)^{2(\delta+1)}$ shows the density reduction due to the redshift, and the exponential term corresponds to $\varphi$'s decay to ultra-relativistic particles (radiation). The radiation energy density is obtained as
\begin{equation}\label{ra}
\rho_r=\frac{4\Gamma M_P^2(\delta+1)^2}{(5+8\delta)t}\left(1-\left(\frac{t}{t_{osc}}\right)^{{-\frac{5}{3}-\frac{8}{3}\delta}}\right).
\end{equation}
Note that for $t\sim \frac{1}{\Gamma}$, we have $\Gamma\sim 3H$ and $\rho_r\sim \rho_{\varphi}$ (for details see (\cite{kolb}). In the above approach we have assumed that during rapid oscillation, the main ingredient of the Universe is the scalar field. But by gradual decay of the inflaton to the radiation, after some time (when $\Gamma\sim 3\gamma H$), the produced  relativistic particles become dominant and compose a fluid in thermal equilibrium. We specify $t_{reh}$ by the time at which $\rho_r(t_{reh})\simeq\rho_{\varphi}(t_{reh})$. The value of the temperature at $t_{reh}$ is denoted by the reheating temperature $T_{reh}$, similar notation will be employed for other parameters at $t_{reh}$. At $t_{reh}$, we have $H^2=\simeq \frac{1}{3M_P^2}\rho_{reh}$. The radiation density and the temperature are related by $\rho_r={g\over 30}\pi^2T^4$, where $g$ is the number of relativistic degrees of freedom, therefore \cite{kolb, Mielczarek1}
\begin{equation}
\Gamma^2\simeq \frac{\pi^2}{10 M_P^2}g_{reh} T_{reh}^4.
\end{equation}

During the rapid oscillation until the radiation dominance, i.e. from $t_{end}$ until $t_{reh}$, the main contribution in the energy density is coming from the scalar field, and $\Gamma \precsim 3H$, hence from (\ref{41}) we derive
\begin{equation}\label{46}
\mathcal{N}_{2}=\ln\bigg({a_{reh}\over a_{end}}\bigg)=-{1\over 3\gamma}\ln\bigg({\rho_{reh}\over \rho_{end}}\bigg),
\end{equation}
which can be rewritten as
\begin{equation}\label{47}
\mathcal{N}_{2}=-{1\over 3\gamma}\ln\bigg({{g_{reh}\over 30}\pi^2T_{reh}^4\over
{\left({-M_P^2C(\delta+{1\over2})}\right)}^{\left({n\over2+2\delta-n\delta}\right)}
{\left(-{n^2\over1+\delta}\right)}^{\left({n(1+\delta)\over 2+2\delta-n\delta}\right)}\lambda^{\left({2(1+\delta)\over2+2\delta-n\delta}\right)}}\bigg),
\end{equation}
where $g_{reh}$ is the number of relativistic degrees of freedom at $t_{reh}$, and we have used (\ref{38.1}).
Using(\ref{34}), we can obtain the scalar field at the horizon crossing
\begin{equation}\label{47.1}
\varphi_\star\approx\bigg({2n(2+2\delta+n-n\delta)\over(1+\delta)(1-n_s)}\bigg)^{\big({1+\delta\over2+2\delta-n\delta}\big)}
(-M_P^2C(\delta+{1\over2})\lambda^\delta)^{\big({1\over2+2\delta-n\delta}\big)}.
\end{equation}
From (\ref{36}) $\lambda$ may be derived as
\begin{eqnarray}\label{47.2}
\lambda\approx (-1)^{n\over2}{(12\pi^2\mathcal{P}_s\sqrt{1+2\delta})^{\big({2\delta+2-n\delta\over2}\big)}
(1+\delta)^{n\over2}\over\big({2(2+2\delta+n-n\delta)\over n(1-n_s)}\big)^{\big({n-n\delta+2\delta+2\over2}\big)}(-(\delta+{1\over2}))^{\big({n+(1+2\delta)(2+2\delta-n\delta)\over-2(1+\delta)}\big)}}\nonumber\\
\times C^{({2-n\over2})}M_P^{(2\delta+4-n\delta-n)}.
\end{eqnarray}
By inserting this relation in equation (\ref{38.1}) $\rho_{end} $ becomes
\begin{eqnarray}\label{47.3}
\rho_{end}\approx
{(12\pi^2\mathcal{P}_s\sqrt{1+2\delta})^{(\delta+1)}(-(\delta+{1\over2}))^{-\big(n+1+2\delta+{n\over n\delta-2\delta-2}\big)}\over n^{\big({(\delta+1)(3n-n\delta+2\delta+2)\over n\delta-2\delta-2}\big)}\big({2(2+2\delta+n-n\delta)\over (1-n_s)}\big)^{\big({(n-n\delta+2\delta+2)(\delta+1)\over2+2\delta-n\delta}\big)}} \nonumber\\
\times CM_P^{2(2+\delta)},
\end{eqnarray}
and finally
\begin{equation}\label{47.4}
\mathcal{N}_{2}=-{1\over 3\gamma}\ln{{{g_{reh}\over 30}\pi^2T_{reh}^4 n^{\big({(\delta+1)(3n-n\delta+2\delta+2)\over n\delta-2\delta-2}\big)}\left({2(2+2\delta+n-n\delta)\over {1-n_s}}\right)^{{(n-n\delta+2\delta+2)(\delta+1)\over2+2\delta-n\delta}}
}\over{(12\pi^2\mathcal{P}_s \sqrt{1+2\delta})^{(\delta+1)}}\left(-(\delta+{1\over2})\right)^{\big(n+1+2\delta+{n\over n\delta-2\delta-2}\big)}CM_P^{2(2+\delta)}}.
\end{equation}

Our above analysis about reheating was based on the simple original studies of perturbative reheating after inflation. There are some problems with this simple model, limiting the range of its applicability, such as collective effects like Bose condensation which alters the decay rate \cite{kof}. The decay rate $\Gamma_\chi\equiv\Gamma_{\varphi\to \chi \chi}=\frac{\sigma^2}{8\pi m}$, derived from the aforementioned three legged interaction, changes when the phase space of bosonic $\chi$ particles is occupied by previously produced bosons. In this situation, we have $\Gamma_{eff.}\simeq \Gamma_\chi(1+2n_k)$, where $n_k$ is the occupation numbers of $\chi$  particles with momentum $\vec{k}$ and $-\vec{k}$: $n_{\vec{k}}=n_{-\vec{k}}=n_k$. For large occupation number, this enhances significantly the decay rate. To get an estimation, by neglecting for a moment the Universe expansion, one finds the simple expression $n_{\chi}\propto e^{\frac{\pi \sigma\Phi t}{2m}}$. This can be derived more precisely from the Mathieu equation corresponding to the equation of motion of the modes of the bosonic field $\chi$ \cite{kof,Mukh}.  Note that the Universe expansion, the back-reaction, and re-scattering of created particles reduce the decay rate enhancement, so the effect of Bose condensation is
actually less than the naive aforementioned estimation \cite{loz}.

If the coupling constants or the inflaton amplitude become large, the perturbative method fails, and higher order Feynman diagrams become relevant. In this situation, the main role in the production of particles is due to the parametric resonance in the preheating era, leading to explosive particles production \cite{Mukh,loz,DM}. This effect must be studied non-perturbatively.  After the inflation, the produced matter field evolves from an initial vacuum state in the background of the oscillating inflaton field. A result of this oscillating background, is a time dependent frequency for the bosonic field ($\chi$) which satisfies the Hill's equation \cite{loz}. Following  Floquet analysis, this may result in a broad parametric resonance and a quick growth of matter in the background of the oscillating inflaton \cite{Mukh,loz,DM}.  By defining  $q$ as $q=\frac{4\sigma \Phi}{m^2}$, one can show that the broad resonance occurs for $q\gtrsim 1$ and we have $n_k\sim e^{2\mu_kmt}$,  where $n_k$ is the occupation number for the bosonic mode $k$, and $\mu_k\sim \mathcal{O}(1)$ is the parameter of instability. For $q\ll  1$ we obtain the narrow resonance  $n_k\sim e^{2\mu_kmt}$,  where $\mu_k\ll 1$ \cite{loz}. The main part of the initial energy is transferred to matter field via the parametric resonance and at the end of broad resonance only a small amount of the initial energy still stored in the inflaton field. Particles created in the preheating era were initially far from thermal equilibrium state, but they reached local thermal equilibrium before BBN. The precise details of reheating era is largely uncertain,  also the realistic picture of preheating faces more complications than the simple aforementioned models.

In our initial perturbative approach, the ultra-relativistic particles are gradually produced (see (\ref{ra})), and until $\rho_r\simeq \rho_\varphi$ ($a=a_{RD}=a_{reh}$) where thermal radiation begins its domination, the Universe is nearly governed by the oscillating inflaton field $\varphi$, whose EoS parameter is given by (\ref{40}). By considering preheating, this assumption fails \cite{DM}. In this situation, as we have explained briefly,  the Universe, besides the oscillating inflaton, is composed of largely produced particles via parametric resonance. Hence we must modify the equation of state (EoS) parameter, and consequently the number of e-folds obtained in (\ref{47.4}). To do so, we employ the method used in \cite{loz}, and
instead of assuming  $w\simeq w_\varphi=\frac{n-2}{n+2}$, we consider an effective equation of state parameter
\begin{equation}
\bar{w}_{eff.}=\frac{\int_{t_{end}}^{t_{reh}}w_{eff.}dt}{t_{reh}-t_{end}}.
\end{equation}
which yields
\begin{equation}\label{47.44}
\mathcal{N}_{2}=-{1\over 3\bar{\gamma}}\ln{{{g_{reh}\over 30}\pi^2T_{reh}^4 n^{\big({(\delta+1)(3n-n\delta+2\delta+2)\over n\delta-2\delta-2}\big)}\left({2(2+2\delta+n-n\delta)\over {1-n_s}}\right)^{{(n-n\delta+2\delta+2)(\delta+1)\over2+2\delta-n\delta}}
}\over{(12\pi^2\mathcal{P}_s \sqrt{1+2\delta})^{(\delta+1)}}\left(-(\delta+{1\over2})\right)^{\big(n+1+2\delta+{n\over n\delta-2\delta-2}\big)}CM_P^{2(2+\delta)}},
\end{equation}
 where $\bar{\gamma}= \bar{w}_{eff.}+1$ (instead of $\gamma= \frac{2n}{n+2}$). We do not know the exact form of $\bar{\gamma}$. But using the fact that after inflation until the thermalization, the Hubble parameter should satisfy $\dot{H}+H^2<0$, we obtain  $\bar{w}_{eff.}>-\frac{1}{3}+\frac{2\delta}{3}$ leading to $\bar{\gamma}>\frac{2}{3}(1+\delta)$.
In the original context where the inflaton gradually decays to particles, we have $\bar{\gamma}\simeq \gamma_\varphi=\frac{2n}{n+2}$ which for the quadratic potential gives $\bar{\gamma}=1$. By considering preheating, we expect that $\bar{\gamma}_{eff.}$ begins from $\bar{\gamma}_{eff.}=1$ and ends to $\bar{\gamma}\simeq \frac{4}{3}$, when the Universe thermalizes. The evolution of EoS between these values depends of the coupling, and also the effective masses of produced particles. This issues has been studied numerically in \cite{DM,pre1}. By considering preheating and an instantaneous thermalization of ultra-relativistic particles, we obtain  $\bar{\gamma}\simeq \frac{4}{3}$ immediately after rapid oscillation \cite{pre1}.

\subsection{Recombination era}

After the coherent oscillation, the Universe contains ultra-relativistic particles in thermal equilibrium and experiences an adiabatic expansion for which \cite{Mielczarek1,kolb}
\begin{equation}\label{49}
{a_{rec}\over a_{reh}}={T_{reh}\over T_{rec}}\left({g_{reh}\over g_{rec}}\right)^{1\over 3}.
\end{equation}
In the recombination era, only photons have relativistic degrees of freedom, therefore  $g_{rec}=2$, and
\begin{equation}\label{50}
\mathcal{N}_{3}= \ln\left({T_{reh}\over
T_{rec}}\left({g_{reh}\over 2}\right)^{1\over 3}\right).
\end{equation}

The temperature redshifts as, $T(z)=T(z=0)(1+z)$, so we can write  $T_{rec}$ in terms of $T_{CMB}$ as
\begin{equation}\label{51}
T_{rec}=(1+z_{rec})T_{CMB}.
\end{equation}
Therefore
\begin{equation}\label{53}
\mathcal{N}_{3}+\mathcal{N}_{4}=\ln\left({T_{reh}\over
T_{CMB}}\left({g_{reh}\over 2}\right)^{1\over3}\right),
\end{equation}
where $\mathcal{N}_{4}$ is the number of e-folds after the recombination era.

\subsection{Reheating Temperature}

To determine the reheating temperature we make use of (\ref{31}). The e-folds from the horizon crossing until the present time, which we take $a_0=1$ , is
$\ln \frac{1}{a_*}$. By using (\ref{23}),(\ref{24}), and (\ref{26}) we obtain
\begin{equation}\label{54}
\ln{1\over a_*}=\ln{H_*\over k_0}\approx \ln\left({2\pi M_P\over k_0}\sqrt{ {n(1-n_s)(1+2\delta)^{-3\over2}\over(2+2\delta+n-n\delta)}\mathcal{P}_s(k_0)}\right).
\end{equation}
By putting this in the left hand side of (\ref{31}), and summing (\ref{35}), (\ref{47.44}) and (\ref{53}) in the right hand side, we derive our main result i.e. $T_{reh}$ as
\begin{equation}\label{55}
T_{reh}\approx  A \pi^{\alpha_1}\left(\frac{M}{M_P}\right)^{\alpha_2} g_{reh}^{\alpha_3}\left({T_{CMB}\over k_0}\right)^{\alpha_4}(1-n_s)^{\alpha_5} \mathcal{P}_s^{\alpha_6} e^{\left(\frac{\alpha_7}{1-n_s}\right)}M_P,
\end{equation}
where we have used $C=M^{-2\delta}$, where $M$ is a mass scale, and $\alpha_n$'s are defined by
\begin{eqnarray}\label{57}
\alpha_1&=&\frac{3\bar{\gamma}-2\delta}{3\bar{\gamma}-4} \nonumber \\
\alpha_2&=&\frac{2\delta}{3\bar{\gamma}-4}\nonumber \\
\alpha_3&=&\frac{-\bar{\gamma}+1}{3\bar{\gamma}-4}\nonumber \\
\alpha_4&=&\frac{3\bar{\gamma}}{3\bar{\gamma}-4}\nonumber \\
\alpha_5&=&-{\frac { \left( n\delta-2\delta-n-2 \right)\left( \delta+1
 \right)}{\left( n\delta-2\delta-2 \right)  \left( 3\bar{\gamma}-4
 \right)}}+3{\frac {\bar{\gamma}}{6\bar{\gamma}-8}}
\nonumber \\
\alpha_6&=&\frac{3\bar{\gamma}-2\delta-2}{6\bar{\gamma}-8} \nonumber \\
\alpha_7&=&3{\frac {\left( -\delta n+2\delta+n+2 \right) \bar{\gamma}}{ \left( -
\delta n+2\,\delta+2 \right)  \left( 3\bar{\gamma}-4 \right) }}
\end{eqnarray}
and the coefficient $A$ is
\begin{eqnarray}\label{56}
A&=&\left(-1 \right) ^{-{\frac {2{\delta}^{2}n+\delta{n}^{2}-4{
\delta}^{2}-n\delta-6\delta-n-2}{3\delta\bar{\gamma} n-6\delta
\bar{\gamma}-4n\delta+8\delta-6\bar{\gamma}+8}}}{2}^{{\frac {{\delta}^{2}n+4
\delta\bar{\gamma} n+\delta{n}^{2}-2{\delta}^{2}-8\delta\bar{\gamma}-
4n\delta-8\bar{\gamma}-2n+2}{ \left( n\delta-2\delta-2 \right)
 \left( 3\bar{\gamma}-4 \right) }}}\nonumber \\
 &&\times{3}^{{\frac {2+\delta}{4-3\bar{\gamma}}}}{
5}^{\frac{1}{4-3\bar{\gamma}}}{n}^{{\frac {-{\delta}^{2}n+3
\delta\bar{\gamma} n+2{\delta}^{2}-6\delta\bar{\gamma}-2n\delta+12
\delta-6\bar{\gamma}+3n+10}{ \left( n\delta-2\delta-2\right)\left(
3\bar{\gamma}-4 \right)}}}\nonumber \\
&&\times \left( 1+2\delta \right) ^{-{\frac {10
{\delta}^{2}n+9\,\delta\bar{\gamma} n+4\delta{n}^{2}-20{\delta}^{
2}-18\delta\bar{\gamma}-2n\delta-32\delta-18\bar{\gamma}-4n-12}{4
 \left( n\delta-2\delta-2 \right)\left( 3\bar{\gamma}-4 \right) }}}\nonumber \\
&&\times
 \left(  \left( -n+2 \right) \delta+n+2 \right) ^{{\frac { \left( 2n
-4\right) {\delta}^{2}+ \left( -8+ \left( -3n+6 \right) \bar{\gamma}
 \right) \delta-2n+6\,\bar{\gamma}-4}{2\, \left( n-2 \right)  \left( 3
\bar{\gamma}-4 \right) \delta-12\bar{\gamma}+16}}}.
\end{eqnarray}

The temperature (\ref{55}), depends on the parameters of the model: $\delta$, $n$, $\bar{\gamma}$, as well as the scale $M$. If we take a quadratic potential, i.e. $n=2$, we obtain
\begin{equation}\label{rel}
\frac{T_{reh,\delta\neq 0}}{T_{reh,\delta=0}}=\beta
(1-n_s)^{\frac{8\delta}{16-12\bar{\gamma}}}\mathcal{P}_s^{\frac{\delta}{4-3\bar{\gamma}}}\left(\frac{M_P}{M}\right)^{\frac{2\delta}{4-3\bar{\gamma}}},
\end{equation}
where  $\beta=\left(-\pi \right) ^{{\frac {2\delta}{4-3\bar{\gamma}}}}{4}^{{\frac {
\delta}{3\bar{\gamma}-4}}}{3}^{{\frac {\delta}{4-3\bar{\gamma}}}} \left( 1+2
\delta \right) ^{{\frac {-10-10\delta-9\,\bar{\gamma}}{12\,\bar{\gamma}-16}}}$. We may obtain a simple expression, if we take $\bar{\gamma}=1$
\begin{equation}\label{rel}
\frac{T_{reh,\delta\neq 0}}{T_{reh,\delta=0}}=\left(\frac{3\pi^2}{4}\right)^\delta (1+2\delta)^{\frac{19+10\delta}{4}}
(1-n_s)^{2\delta}\mathcal{P}_s^{\delta}\left(\frac{M_P}{M}\right)^{2\delta}.
\end{equation}

For $n=2$, $\delta=0$ and $\bar{\gamma}=1$, the reheating temperature (\ref{55})reduces to
\begin{equation}\label{58}
T_{reh}\approx \frac{45}{2\pi^3} \sqrt{{(1-n_s)\over\mathcal{P}_s}}\exp{\Big({6\over1-n_s}\Big)}\bigg({k_0\over T_{CMB}}\bigg)^3M_P,
\end{equation}
which is the temperature obtained in \cite{Mielczarek1} in the context of the standard general relativity.
In this situation, and for $\delta=1$, we have
\begin{equation}\label{60}
{T_{reh,\delta=1}\over T_{reh,\delta=0}}\approx 21306 \mathcal{P}_s(1-n_s)^2{M_P^2\over M^2}.
\end{equation}
By taking $g_{reh}=106.75$,  $k_0=0.05Mpc^{-1}$, and by using $(68\% CL, TT,TE,EE+lowE+lensing)$\cite{planck}
\begin{eqnarray}\label{61}
\ln\Big(10^{10}\mathcal{P}_s(k_0)\Big)&=& 3.044\pm 0.014\nonumber \\
n_s&=&0.9645\pm 0.0042,
\end{eqnarray}
(\ref{60}) reduces to
\begin{equation}\label{62}
{T_{reh,\delta=1}\over T_{reh,\delta=0}}\approx 5.635\times 10^{-8}{M_P^2\over M^2}.
\end{equation}
  So the temperature is much less than $T_{reh,\delta=0}$ unless $M_P\gg M$. For example, by taking  $M=10^{-4}M_P$ we obtain $T_{reh}\approx 5.635T_{reh,\delta=0}\sim 6\times10^{14}GeV$.

\section{Conclusion}
We considered inflation in a modified teleparallel model of gravity (see (\ref{1})), in which a scalar field is responsible to reheat the Universe after the inflationary era. To determine the reheating temperature, we used the cosmological perturbations to find the number of e-folds from the horizon exit of a pilot scale, until now . In addition, we divided the evolution of the Universe into different segments and obtained the corresponding efolds in each segment and summed over them. By equating efolds numbers derived from these two methods, we achieved to obtain an expression for the reheating temperature in terms of the CMB temperature, the spectral index, the power spectrum and the parameters of the model.

\end{document}